\documentstyle[12pt,aasms4]{article}

\begin{document}
\title {Asymmetric Explosions of Type Ia Supernovae}

\author{Cristian R. Ghezzi\altaffilmark{1}, Elisabete M. de Gouveia Dal 
Pino\altaffilmark{1}
\& Jorge E. Horvath\altaffilmark{1} }
\altaffiltext{1}
{Instituto Astron\^omico e Geof\'{\i}sico, University
of S\~ao Paulo, Av. Miguel St\'efano, 4200, S\~ao Paulo
04301-904, SP, Brasil;
E-mail: dalpino@iagusp.usp.br, ghezzi@iagusp.usp.br }

\begin{abstract}

 The burning speed of a thermonuclear supernova front
could be described by the fractal model of combustion.
We have  examined the effects of magnetic fields
on the fractalization of the front considering a white dwarf
with a nearly dipolar magnetic field and found
an intrinsic asymmetry on the velocity field of the expanding plasma. 
For white dwarf's magnetic fields of  $10^8-10^9$ G at the
surface, and assuming a field roughly 10 times greater near the center,
we have found asymmetries in the velocity field 
$> 10-20 \%$ at $\rho \sim 10^8\,{\rm g \,cm}^{-3}$ ,
produced between the magnetic polar and the equatorial 
axis of the remnant. This effect 
may be related to the asphericities 
inferred from spectro-polarimetric observations of very young
SN Ia remnants (for example: the SN 1999by).
In the present work, we analyse the dependence of the asymmetry with the 
composition of the white dwarf progenitor.

\end{abstract}

\section{Introduction}

The explosion of a type Ia supernova begins with the 
combustion
at the center of a Chandrasekhar mass white dwarf of carbon-oxygen 
(C+O)  or 
oxygen-neon-magnesium (O+Ne+Mg) fuels. The heat is transported mainly 
by conduction due to degenerate and completely relativistic electrons 
as a subsonic deflagration wave propagatating outwardly of the star. 
The deflagration front born laminar is subject to several hydrodynamic 
instabilities 
such as Landau-Darrieus (LD) and Rayleigh Taylor (RT) instability
(Arnett \& Livne 1994, Khokhlov 1993) that produce an 
increment of the area at which the nuclear reactions take place.
This causes an increase of the nuclear energy generation rate and 
consequently an acceleration of the front. 

The combustion front is stabilized by the merging of cells, the 
formation of cusps, and  
the expansion of the exploding star. This leads to the formation of a 
cellular 
structure at microscopic scales.  
The bubbles that grow due to RT instability are also subject to 
Kelvin-Helmoltz (KH) or 
shear instability when nonlinear stabilization fails. 
The onset of the KH instability marks the 
transition to the fully developed turbulence regime at the lower scales.
During this, fluid motions are characterized by the formation of a 
turbulent cascade in the inertial scales where viscous 
dissipation is not important. This turbulence can be described 
by the Kolmogorov's scaling law. 

The fractal model for the combustion 
(e.g., Timmes \& Woosley 1992, Niemeyer \& Woosley 1997) 
has achieved some success on describing the acceleration 
of wrinkled flames both in experiments and also in 
numerical simulations. 
The velocity of the flame in this case is given by:

\begin{equation}   
v_{frac} \, = v_{lam} (L/l_{min})^{D-2}
\end{equation}   

\noindent
Where $v_{frac}$ is the effective fractal velocity and
$v_{lam}$  is the laminar velocity of the flame; 
$L$ and $l_{min}$ 
are the greatest and minimum scales, respectively, 
of preturbations which are R-T unstable; 
and D is the fractal dimension of the front.  
The derivation of the value of D for a turbulent combustion regime has given
$D =  2.25 -2.5$ 
(see Ghezzi, de Gouveia Dal Pino \& Horvath 2001), 
which is in agreement with previous numerical results  (Blinnikov, 
Sasorov \& 
Woosley 1995).

\section{Magnetic field effects on the fractalization of the combustion 
front}

We have incorporated the effects of magnetic fields on the 
fractal growth of the
combustion front assuming  that the 
progenitor star of a SN Ia is a magnetized white dwarf with a centered 
dipolar magnetic field with strengths at the surface $~10^8 - 10^9$ G (see 
Jordan 1992) and roughly 10 times greater near the center. Using the 
fractal model described above we have found that the presence of the 
magnetic field causes the effective velocity of the flame at the magnetic  
poles ($v_{pol}$) to be larger than that at  the equator of the star 
thus producing an asymmetry in the explosion of the SN Ia (Ghezzi, de 
Gouveia Dal Pino \& Horvath 2001):
   
\begin{equation}
\label{asym}
\frac{v_{pol}}{v_{eq}}=
\biggl( \frac{B^{2}/8\pi + \rho \, v_{lam}^{2}/2}{\rho \, 
v_{lam}^{2}/2} \biggr) ^{D-2} .
\end{equation}

For the evaluation 
of this equation 
we have used a fractal dimension $D=2.5$, 
and taken the remaining
data from Timmes \& Woosley (1992).
We note that our analysis is applicable only for densities $\rho \geq 
10^{7}\,{\rm g\,cm^{-3}}$, 
since the turbulent motions could
 destroy the corrugated flamelet 
regime
for lower densities (Niemeyer \& Woosley 1997), and in this case the fractal model
is no longer aplicable.

 Figure 1 displays the percentage of asymmetry in the
velocity field as a function of the internal 
magnetic field strength  at a 
a middle radial distance $\sim 8 \times 10^{7}\,{\rm 
cm}$ from the center of the star (where the
fuel density is $\rho\sim10^{9}\,{\rm g\,cm^{-3}}$)  
for white dwarfs with 
different
compositions. For example, 
for a composition of $X(^{12}C)=0.2$ and $X(^{16}O)=0.8$, 
we have used
 $v_{lam}=0.415\times 10^{5} \, {\rm cm\,s^{-1}}$.  
We see that the asymmetry
is  sensitive to the composition of the progenitor.
Heavier progenitors show higher asymmetries since their 
laminar velocities are smaller, so that the kinetic energy term becomes less dominant 
with respect to the magnetic energy term in the asymmetry equation 
above.

Figure 2 shows the  asymmetry on the combustion front 
at a density $\rho = 10^8\,\, {\rm g\,cm}^{-3}$. The total asymmetry 
increases with increasing radius (or 
decreasing density) and  
is larger for progenitors 
with heavier compositions.

\section{Conclusions }

An asymmetry in the velocity field is developed by the presence of a
dipolar magnetic field during the fractal growth of the deflagration front of a 
type Ia supernova  that can lead to the formation of a prolate remnant. 
The magnetic field introduces an effective surface tension 
in the equator of the white dwarf progenitor that reduces the velocity
of the combustion front at the equator, $v_{eq}$, with respect to the 
velocity at the poles,
$v_{pol}$, so that $v_{pol} > v_{eq}$. The asymmetry is larger for 
heavier progenitors\footnote{In this work we calculated the "instantaneous asymmetry values" at a 
given density, as we will show (see Ghezzi et al. 2002, and Ghezzi 2002) integrating the effect over the explosion
leads to higher asphericity of the remnant.}.
In particular, for progenitors with a composition 
$X(^{12}{\rm C})=0.2\,\,X(^{16}{\rm O})=0.8\,$, $\Delta \rho /\rho=0.415$, and 
surface magnetic fields $\sim 10^{8}$ G, a 
$10\,$ to $20\,\%$ asymmetry has been found at a middle distance from the center of 
the star (see Fig. 2). 
                       
 As only a small fraction of the observed white dwarfs are inferred to 
have magnetic fields higher than
about $10^{8}$ G, asymmetries
are not expected to occur very frequently.
Nonetheless, recent spectropolarimetric observations have revealed a 
linear polarization
component in the radiation of very young SN Ia remnants,  
which suggests that prolate atmospheres with asymmetries $> 17\,\%$ 
are producing it (see Leonard, Filippenko, \& Matheson 1999, 
Wang , Wheeler \& H\"oflich 1997 and Howell, Hoeflich, Wang \& Wheeler 2001).
The model presented here offers a plausible explanation for such 
observations.

\acknowledgements
This paper has been partially supported by grants of the Brazilian
Agencies FAPESP and CNPq, and by the PRONEX project.

\newpage
{\bf Figure Caption}

Figure 1. Asymmetry percentage in the field velocity for progenitors 
with different initial compositions
at $\rho \simeq 10^{9}$ ${\rm g\,\,cm^{-3}}$, as a function of the 
central magnetic field strength. 

 The asymmetry has lower values at higher densities. So this figure represent
the lower values for the asymmetry, since the central density values of the  
progenitors are of the order of $\sim 10^{9}$ ${\rm g\,\,cm^{-3}}$. From this figure
is possible to see that heavier compositions show higher asymmetry values.

Figure 2. Asymmetry percentage  for two different progenitors with 
compositions 
$X(^{12}{\rm C})=0.2$ $X(^{16}{\rm O})=0.8$, 
for the full line and
$X(^{12}{\rm C})=0.5 $ $X(^{16}{\rm O})=0.5$, 
for the dotted line. The calculation were made at a density of
$\rho = 10^8\,{\rm g\,cm}^{-3}$.


\begin{thebibliography}{}

\bibitem{}  Arnett W. D., \& Livne E. 1994a, ApJ, 427, 315

\bibitem{}  Blinnikov S. I., Sasorov P. V. \& 
Woosley S. E., Space Science Review, v 74,  299, 1995


\bibitem{} Ghezzi, C. R., Gouveia Dal Pino, E. M. \& 
Horvath, J. E. 2001, ApJ Letters, 548, L193

\bibitem{}  Ghezzi, C. R., Gouveia Dal Pino, E. M. \&
Horvath, J. E., 2002, in preparation

\bibitem{} Ghezzi, C. R., PhD thesis, Institute of Astronomy and Geophysics of the S\~ao Paulo University, 
S\~ao Paulo,  Brasil, May 2002

\bibitem{} Howell, D. A., Hoeflich, P., Wang, L., \& Wheeler, J. C., astro-ph/0101520

\bibitem{} Jordan S., Astron. Astrophys., 265, 570, 1992

\bibitem{} Khokhlov A., ApJ, 419, L77, 1993 


\bibitem{} Niemeyer J. C. \& Woosley, S. E., 1997, ApJ, 475, 74


\bibitem{} Timmes F. X. \& Woosley S. E., ApJ, 396, 667,
1992 

\end{thebibliography}
\end{document}